\documentclass{article}
\usepackage{spconf,amsmath,amssymb,graphicx}
\usepackage{url}
\usepackage{booktabs}
\usepackage{multirow}
\usepackage{subcaption}

\newcommand{\topk}{\texttt{TopK~}}

\title{Sparse deepfake detection promotes better disentanglement}
\name{Antoine Teissier, Marie Tahon, Nicolas Dugué, Aghilas Sini.}
\address{LIUM, Le Mans University}
\begin{document}
%\ninept
%
\maketitle
\begin{abstract}
Due to the rapid progress of speech synthesis, deepfake detection has become a major concern in the speech processing community. Because it is a critical task, systems must not only be efficient and robust, but also provide interpretable explanations.
Among the different approaches for explainability, we focus on the interpretation of latent representations. In such paper, we focus on the last layer of embeddings of AASIST, a deepfake detection architecture. We use a \topk activation inspired by SAEs on this layer to obtain sparse representations which are used in the decision process.
We demonstrate that sparse deepfake detection can improve detection performance, with an EER of 23.36\% on ASVSpoof5 test set, with 95\% of sparsity.
We then show that these representations provide better disentanglement, using completeness and modularity metrics based on mutual information. Notably, some attacks are directly encoded in the latent space.
\end{abstract}
\section{Introduction}
\label{sec:intro}

Deepfake detection is a challenging task that received significant attention due to the rapid progress of modern speech synthesis systems. As a critical task, systems must be performant and, ideally, provide interpretable explanations for its decisions to inspire user confidence. Performance of such systems is influenced by multiple factors, including recording conditions, speaker identity, and the nature of the attacks. Furthermore, attacks evolve with improvements in speech synthesis systems, making it difficult to anticipate new attack types at inference time while keeping explainability.

There is currently an ongoing effort of the scientific community to explainable AI (XAI), especially under the neural paradigm.
For deepfake detection, a seminal study proposed to use Shapley values to highlight the parts of the signals that may be responsible for a given decision \cite{ge_explaining_2022}. 
More generally, one of the line of research is to explain the role of the components of the system in the decision, such as attention analysis~\cite{wiegreffe-pinter-2019-attention} or mechanistic interpretability~\cite{kulkarni2025mechanistic}.

We focus on interpreting latent representations of neural architectures. Probing can reveal human-understandable features~\cite{deseyssel2022probing, liu2024explaining}, but they are spread across dimensions.
We thus advocate for a more disentangled latent space for the sake of interpretability. 
To do so, we need i) to design new architectures that favor better explanations, and ii) to develop protocols to analyze which factors are encoded in the latent space.

Different approaches have been investigated to train models where latent representations satisfy some constraints regarding interpretability, such as sparsity~\cite{murphy2012learning} or orthogonality~\cite{almudevar2024unsupervised}. These constraints are supposed to promote better disentanglement of the latent space towards generative factors. 
Recently, many approaches have been proposed to auto-encode sparse latent spaces (SAE) with \topk L0 ~\cite{gao2024scaling}, L1 sparsity~\cite{subramanian2018spine} or orthogonality~\cite{piaggesi2024dine} losses.
%More precisely, \topk loss only keep the $k$ most activated dimensions, zeroing the rest.
% J'ai ajouté l'info topK L0, qui résume cette idée. Et on la réexplicite ensuite. Donc je supprime pouir gagner de la place

In the literature, disentangled metrics are often based on two principles: \textit{modularity} (one dimension relates to at most one factor) and \textit{completeness} (one factor relates to at most one dimension)~\cite{carbonneau2022measuring}.
However, these principles rely on the mutual independence of factors, which might be a strong issue in speech processing as many factors are related ($F_0$ and gender for example)~\cite{almudevar2024definingmeasuring}.
Therefore, choosing a set of factors remains a difficult task~\cite{fucci-etal-2024-explainability}.
Many others define their set of factors among speech production parameters such as prosody, phonemes, or even spectrograms~\cite{pasad2023comparative}.
In this paper, we use attacks as factors, as in~\cite{MISHRA2026101840}.

%Related work has explored these challenges from different angles. Recent speech synthesis systems~\cite{li2023styletts} have begun to integrate deepfake detection mechanisms into their training processes, with the goal of improving the naturalness and robustness of generated speech. Conversely, the speech synthesis community has also started to leverage deepfake detection for evaluation purposes, for example by adapting detection-based predictors as surrogates for human quality assessments such as MOS~\cite{miniconi2025using}. Beyond detection, interpretability has been investigated through approaches that attempt to attribute decisions to meaningful factors~\cite{ge_explaining_2022}, as well as through techniques based on sparsity, such as sparse autoencoders or \topk feature selection.

%blabla sur metriques de démelage

In this work, we employ AASIST~\cite{jung2022aasist}, a well-known graph-based binary classifier, adding a \topk activation on the last hidden layer to train sparse deepfake detection models.
We then evaluate how this activation impact performance and disentanglement, considering different sizes for the last hidden layer. To evaluate performance, we consider data from ASV spoof challenge, and EER results. Regarding disentanglement, we follow previous work and consider the set of attacks as factors. We analyze the sparse latent space using mutual information with two disentanglement metrics -- completeness and modularity-- with the goal of distinguishing dimensions that primarily encode attack-related information.
%extract the most prominent features from the classifier’s bottleneck representations. 
%plausible factors that can provide sustainable explanations for detection decisions without compromising model performance. 
%To achieve this, we design several experiments to explore how \topk affects detection performance and structures the latent space in a more disentangled manner. 
Our contributions are threefolds: (i) the first use of sparse constraints for deepfake detection; (ii) evidences that \topk achieve both performance and disentanglement; (iii) evidences that attacks are directly encoded in the latent space.

\section{Proposed Approach}
\label{sec:general}

\subsection{Dataset}

Models are trained and evaluated on the dataset proposed during the Automatic Speaker Verification Spoofing and Countermeasures (ASVspoof) challenge in 2024~\cite{wang24_asvspoof}. The data comes from the English part of MLS~\cite{pratap20_interspeech}, a read speech (audiobook) corpus, with numerous speakers recorded in various acoustic environments.

Attacks have been designed to be more challenging than the previous editions. Spoof data was generated with recent Text-To-Speech (TTS) or Voice Conversion (VC) systems, but also by using codecs on both bona fide and spoof data.
One main difficulty is that the attacks are different in each partition (train, development, or test). Moreover, the attacks in the test set are more diverse than in the train and dev sets as shown in Table~\ref{tab:Distribution_du_dataset_ASVspoof5}. Consequently, there is a huge risk of overfitting.

We would like to highlight two difficult attacks. First, the unit-selection A12 is composed of a concatenation of bona fide small signals. Artifacts might occur when the junction is not well suited.
A subset of TTS and VC attacks are combined with Malafide (A18, A20, A23 A30) and Malacupola (A27, A30, A31, A32) filtering, however this specific process is done only in the test set.
Malafide~\cite{panariello23b_interspeech} introduces a convolutional noise using an optimised
linear time-invariant filter, while Malacupola~\cite{todisco24_asvspoof} modifies amplitude, phase, and frequency content in non-linear fashion. Both filters were designed to fool countermeasure systems.

\begin{table}[t!]
\caption{ASVspoof5 dataset characteristics, number of attacks and speakers}
    \centering
        \begin{tabular}{|c|c|c|c|c|c|}
            \toprule
            Data & \#Samples & Spoof& Bonafide & \#Atts. & \# Spks. \\
             & & (\%)& (\%) &  &  \\
            \midrule
            train& 182,357 & 89.3 & 10.3 & 8 & 400 \\
            dev& 140,950 & 77.8 & 22.2 & 8 & 785 \\
            test& 680,774 & 79.6 & 20.4 & 17 & 737  \\
%            ASVspoof5\_train\_lite& 54,707 & 89.6 & 10.4 & 8 & 400 & 50.3 & 49.7 \\
%            ASVspoof5\_dev\_lite& 42,285 & 77.8 & 22.2 & 8 & 753 & 51.1 & 48.9 \\
\bottomrule
        \end{tabular}
    
    \label{tab:Distribution_du_dataset_ASVspoof5}
\end{table}

Performances of the deepfake detection model are evaluated with Equal Error Rate (EER) and Detection Cost Function (DCF) on the \textit{bonafide} class.

\subsection{Proposed sparse model}

AASIST~\cite{jung2022aasist} is a graph-based classifier designed for anti-spoofing. This architecture achieves competitive performance with only 0.3M parameters, compared to state-of-the-art models averaging 300M. %The implementation is open-source\footnote{\url{https://github.com/asvspoof-challenge/asvspoof5/tree/main/Baseline-AASIST}} and 
%It was trained from scratch in all configurations, without any pretrained feature extractors, to ensure fair and interpretable results.
It takes raw audio samples as input. In the first stage, it extracts general features, while in the second stage it focuses on spectro-temporal patterns using graph-based representations, followed by an attention mechanism and a pooling process that provides latent representations. The final component is a binary classifier that outputs a bona fide/spoofed pseudo-probability.

\textbf{\topk} activation is applied on the last hidden layer of AASIST model to ensure a given level of sparsity -- the $k$ highest values are retained, while the rest is set to 0 -- in the weights of the layer during the training process. Such an activation enforces sparsity at the vector level, by allowing a maximum of $k$ non-zero values in each embedding vector~\cite{\cite{gao2024scaling}}. %The regularization strategy is enforced through a specific \topk activation function applied after the sigmoid $\sigma$ in $\phi = TopK(\sigma(f(\mathbf{x}_i)))$ as described in \cite{gao2024scaling}. It keeps the $k$ most activated values per encoded vector, zeroing the rest.
To better enforce sparsity, one might also increase the size of the vector~\cite{subramanian2018spine}. As the size of the last layer of AASIST is 160, we consider $D=160$ and $320$ dimensions for this last layer. In order to investigate the effect of sparsity on both the performances for deepfake detection, and the disentanglement, we consider several values of $k$: 20, 50 and 100.

\section{Deepfake detection results}

\subsection{Performance results}

Deepfake detection results are presented in Table~\ref{tab:results160dev}. The baseline model with is indeed the original AASIST.
We observe that with $D=160$, decreasing $k$ induces a degradation of the performance both in terms of EER and DCF on the development set.
However, we observe on the test set that more sparsity implies a better generalization, as all EER and DCF decrease with $k$.
As a reminder, ASVSpoof5 test set is bigger and more diverse than the train and development sets.
The results obtained with $D=320$ and no sparsity are lower than those obtained with the original size 160 on the dev set, but better on the test set.
When introducing sparsity, $D=320$ allows better performances in almost all cases for on both dev and test sets.
Especially, on the dev, we notice that $k=50$ gives the best performances. And the model that obtains the best performances of all on the test set is obtained with $k=20$, with roughly $95\%$ of sparsity.

\begin{table}[t!]
\centering
\caption{Results on ASVspoof5 according to $k$, best epoch.}
\begin{tabular}{|c|c|c|c|c|c|c|c|c|}
\toprule
\multirow{2}{*}{$D$} & & & \multicolumn{2}{c|}{dev set} & \multicolumn{2}{c|}{test set} \\
&\textbf{$k$} & \textbf{epoch} & \textbf{EER} & \textbf{DCF} & \textbf{EER} & \textbf{DCF}\\
    \midrule
\multirow{4}{*}{160}    &- & 10 & \textbf{15.44} & \textbf{0.3166} & 28.94 & 0.6725 \\
    \cline{2-7}
    &100 & 7 & 18.81 & 0.3739 & \textbf{25.27} & \textbf{0.5957} \\
    &50 & 6 & 17.27 & 0.3403 & 25.86 & 0.6107 \\
    &20 & 5 & 19.82 & 0.3997 & 26.96 & 0.6669 \\
    \bottomrule
\multirow{4}{*}{320}  &    - &  12 &  17.18 &  0.3489 & 25.88 & 0.6105   \\
    \cline{2-7}
    &100 & 12  & 17.33 & 0.3713  & 31.96 & 0.7187 \\
    & 50 &  9 &  \textbf{16.44} &  \textbf{0.3473} & 25.57 & 0.6167 \\
    &20 & 18 & 17.18 & 0.3489 & \textbf{23.36 }& \textbf{0.5807} \\
    \bottomrule
\end{tabular}

\label{tab:results160dev}
\end{table}

With such results, we demonstrate the efficiency of the \topk activation combined with oversized hidden layer for deepfake classification. In particular, we show that the sparsest model with $95\%$ of zeros obtained the best results.

 %As we have around 20 factors (attacks), we might not reach good disentanglement between dimensions and factors.
%Therefore, we decided that $k=50$ is the best compromise, to reach good disentanglement.

\subsection{Performance analysis per attack}

We also calculate the performance obtained for each attack of the dev and test sets. Because of space constraints, the results are not reported here.
First, we observe that the attack A12 reach a poor performance on dev set (EER $\simeq 0.8$) for both the baseline model ($k=160$) and the model trained with \topk ($k=50$). All other attacks get an EER below $0.2$.
%We notice that the attack A12 reached a poor EER (over 0.8 on the dev set) in comparison to the other attacks.
Indeed, as discussed before, A12 corresponds to the unit-selection attack, which explains this bad result.
Therefore, we will discard this attack in the following analysis regarding disentanglement. Indeed, an attack that is not detected correctly by the system is not encoded in a disentangled subspace.
Following the same line, we decided to remove all attacks of the test set for which the performance is $EER > 0.2$.
We then remove attacks derived from A12 (A20), the one which includes Malacopula (A27, A30, A31 and A32) or Malafide (A18, A20), as well as A19, A25, and A26.

Still, we  notice that the addition of the \topk activation tends to improve performances on attacks that were not well detected with the baseline model.

%\begin{adjustwidth}{-1.5cm}{-1.5cm}
%\begin{figure*}[htb]
%\begin{minipage}[b]{0.48\linewidth}
%  \centering
%  \centerline{\includegraphics[width=8.5cm]{img/tauxEERDev.png}}
%  \vspace{2.0cm}
%  \centerline{(a) Dev set}\medskip
%\end{minipage}
%\hfill
%\begin{minipage}[b]{0.48\linewidth}
%  \centering
%  \centerline{\includegraphics[width=8.5cm]{img/tauxEERTest.png}}
%  \vspace{2.0cm}
%  \centerline{(b) Test set}\medskip
%\end{minipage}
%  \caption{Comparison of EER performance with layer size of $D=160$}
%  \label{fig:compEER}
%\end{figure*}
%\end{adjustwidth}
\begin{table*}[h!]
    \centering
    \caption{Average completeness on 4 seeds regarding 7 attacks and bonafide as factors on ASVspoof5.}
    %\resizebox{0.9\textwidth}{!}{
    \begin{tabular}{|c|c|c|c|c|c|c|c|c|c|c|c|c|}
    \toprule
    &\textbf{$D$} & \textbf{$k$} & \textbf{A09} & \textbf{A10} & \textbf{A11} & \textbf{A13} & \textbf{A14} & \textbf{A15} & \textbf{A16} & \textbf{Bonafide} \\
    \midrule
\multirow{4}{*}{Dev set}    &160 & - &0.02&	0.02&	0.02&	0.05&	0.02	&0.02&	0.01&	0.02\\
    &160 & 50 & 0.11&	0.12&	0.09&	0.12	&0.08&	0.10&	0.10&	0.11\\
    &320 & - & 0.11&	0.06&	0.05	&0.09&	0.06	&0.05&	0.05&	0.06\\
    &\textbf{320} & \textbf{50} & \textbf{0.21}	& \textbf{0.22} &	\textbf{0.21}&	\textbf{0.20}&	\textbf{0.20}	&\textbf{0.19}	&\textbf{0.19}&	\textbf{0.24}\\
    \toprule
        & & & \textbf{A17} & \textbf{A21} & \textbf{A22} & \textbf{A23} & \textbf{A24} & \textbf{A28} & \textbf{A29} & \textbf{Bonafide} \\
    \midrule
\multirow{4}{*}{Test set}       & 160 & - & 0.02&	0.03&	0.03&	0.02&	0.02&	0.02&	0.02&	0.03\\
    &160 & 50&0.11	&0.11&	0.11&	0.10&	0.07&	0.09&	0.06&	0.12\\
    &320& -&0.06	&0.06&	0.06	&0.05&	0.05&	0.05&	0.05&	0.06\\
    &\textbf{320} & \textbf{50} &\textbf{0.23}&	\textbf{0.23}&	\textbf{0.22}&	\textbf{0.22}	&\textbf{0.18}&	\textbf{0.21}&	\textbf{0.18}&	\textbf{0.24}\\
    \bottomrule
    \end{tabular}
    %}

    \label{tab:resultsmidci_comp_dev_att}
\end{table*}

\section{Disentanglement towards attacks}

In this section, we will follow the \textit{completeness} and \textit{modularity} definitions and metrics proposed in recent articles~\cite{carbonneau2022measuring,zhang2023extension}.
To calculate disentanglement metrics, we refer to a set of $F$ factors which corresponds to the binary presence of a given attack.
Most of the analysis is reported on the dev set for the sake of space, but has been done on the test set as well, leading to the same conclusions.

We report completeness and modularity metrics obtained with deepfake detection models with size of the last layer equal to $D=160, 320$.
In order to assess the effective use of the \topk loss to address better disentanglement, we also evaluate with $k=D$ or $k=50$ as this value has shown good performance in terms of deepfake classification.
Moreover, we trained 4 different models for each configuration with different seeds. This results in the training of 16 models. Completeness and modularity metrics are averaged over the 4 seeds for each configuration.

\begin{figure}[h]
    \centering
    \includegraphics[width=1.0\linewidth]{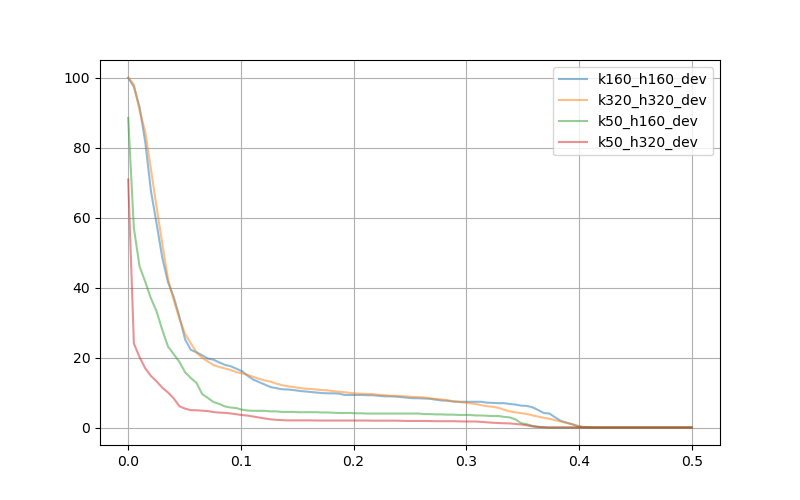}
    \caption{In ordinate, number of values in \% above the nMI value in abscissa. For ex: 20\% of the values are above 0.05 with $D=160$ and $k=50$ (green line).}
    \label{fig:miNonNulDev}
\end{figure}

\subsection{Mutual Information}

We use the mutual information (MI) calculated between each dimension and each factor to define the importance matrix. This matrix $\mathbf{M} \in \mathbb{R}^{D\times F}$ encodes how well a factor $f$ is encoded with a dimension, and inversely. In our case, all factors are independent.
In our case, $D=160$ or $360$, according to the size of the last layer, and the number of factors used for the analysis is the 8 attacks for which the model gives low errors ($EER < 0.2$). 
%In the end 8 attacks have been selected for the analysis.

We first normalize the mutual entropy $\mathbf{M}_{d,f} = \frac{2 MI(d, f)}{H(d) + H(f)}$ with the entropy of the factor $f_i$ and the entropy of the dimension to limit scale effects.

Figure~\ref{fig:miNonNulDev} displays the number values in $\mathbf{M}$ above a given threshold. It gives an overview of how are distributed the values of the nMI. With \topk activations (green and red lines), we can see that the distribution is more skewed towards zero. There are way more zeroes in the matrix, without any loss of performances, indicating that information regarding the attacks is less dispersed across dimensions. It seems that sparsity in the latent representations leads to sparse MI matrix, which might also indicate disentanglement.

\subsection{Completeness}

\textit{Completeness} is satisfied when a factor is represented by a single dimension, a given dimension can encode several factors. In our case, it would mean that an attack is encoded by a single dimension.
Completeness is defined with eq.~\ref{eq:dcicomp}, where $p(f| f) = M_{d, f} / \sum_f M_{d, f}$ is the probability of a dimension $d$ conditioned by factor $f$.

\begin{equation}
    \label{eq:dcicomp}
    \text{comp}(f)= 1 + \sum_d p(f|d) \log_D p(f|d)
\end{equation}

From Table~\ref{tab:resultsmidci_comp_dev_att}, we observe that for each size $D$, the \topk activation with $k=50$ favors completeness.
Moreover, with higher $D$, the completeness is even better.
Thus, we confirm that increasing the latent space and constraining sparsity promotes a better disentanglement.
We hypothesize that with such constraints, the dimensions are more atomic, in the sense that they represent more atomic \textit{i.e.} low level features.

\subsection{Modularity}

\textit{Modularity} is achieved when a factor affects only a subspace of the representation space. In other words, high modularity implies that a given dimension encodes only few factors, but multiple dimensions can represent the same factor. High modularity would be achieved when a dimension encodes only one or few attacks.
Modularity is defined with eq.~\ref{eq:dcimod}, where $p(d| f) = M_{d, f} / \sum_d M_{d, f}$ is the probability of a dimension $d$ conditioned by factor $f$.

\begin{equation}
    \label{eq:dcimod}
    \text{mod}(d)= 1 + \sum_{f\in \mathcal{F}} p(d|f) \log_F p(d|f) 
\end{equation}

\begin{figure}[htbp]
  \centering
  %\begin{minipage}{0.48\linewidth}
    \centering
    \includegraphics[width=\linewidth]{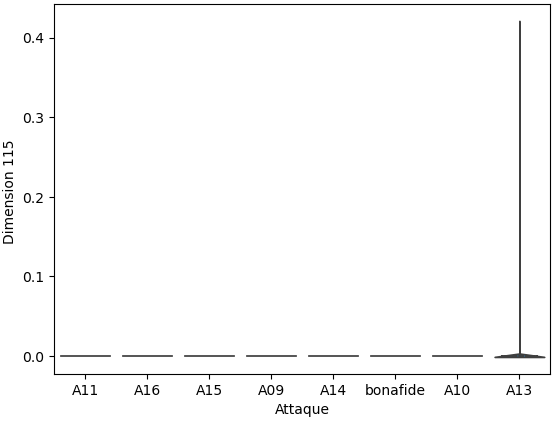}
    \centerline{(a) Activated attacks for dimension 115}\medskip
    \label{fig:dim115_att}
  %\end{minipage}
  %\hfill
  %\begin{minipage}{0.49\linewidth}
   
    \includegraphics[width=\linewidth]{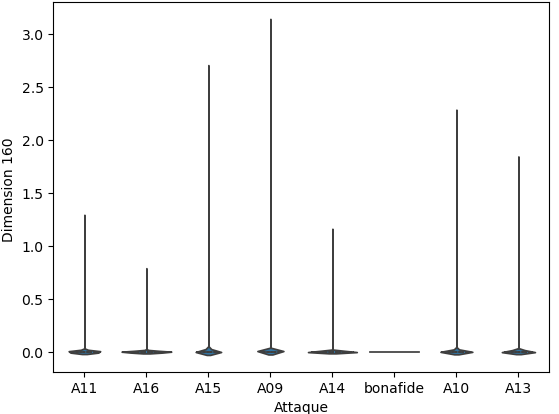}
    \centerline{(b) Activated attacks for dimension 160}\medskip
    \label{fig:dim160_att}
  %\end{minipage}
  \caption{nMI per attack for two dimensions, one with high modularity and the other with medium (Dev set).}
  \label{fig:dim_att}
\end{figure}

We do not report all modularity values for each model for better readability. 
%Figure~\ref{fig:modulariteDev_att} shows modularity values by dimension for the configuration with the best completeness ($D=320$, $k=50$).
%Dimensions with a high modularity are supposed to encode very few factors. 
Instead, we cherry-picked two dimensions, one with the highest modularity ($\text{mod}(115) = 1.00$), the other with medium modularity ($\text{mod}(160) = 0.74$).
For these two dimensions, we plot the distribution of the nMI for each attack of the dev set in Figure~\ref{fig:dim_att}.
On the one hand, we notice that the dimension 115 with the highest modularity indeed encodes only one attack (A13). 
On the other hand, the dimension 160 with a medium modularity encodes all attacks but bona fide.

%\begin{figure}[h!]
    %\centering
    %\includegraphics[width=1.0\linewidth]{img/modulariteDev_att.png}
    %\caption{Modularity values according to attacks for one model with $D=320$ and $k=50$. Results obtained on the Dev.}
    %\label{fig:modulariteDev_att}
%\end{figure}

\section{Conclusion}

This work investigates deepfake detection from an interpretability point of view.
We propose a novel approach to train sparse deepfake detectors with the addition of a \topk activation on the last hidden layer of AASIST model.
We demonstrate that our simple approach improves detection results with an EER of 23.36\% with $k=20$ and when the size of the last hidden layer is large enough ($D=320$ in our case). This result is very interesting, as the sparsity of the latent space reaches 95\% of zeros.
We also show that \topk promotes disentanglement, as it improves both completeness and modularity when the considered factors are attacks.
Finally, a deeper analysis of the mutual information between dimensions and attacks shows that dimensions with high modularity can encode a unique attack.

The proposed approach shows a promising path towards more complex interpretability analysis. In the future, we would like to not only disentangle the latent space towards attacks, but also towards more fine-grained attributes of the spoofed or bona fide signals.

\newpage
\bibliographystyle{IEEEbib}
\bibliography{strings,refs}

\end{document}